\documentclass[
 aps,
 prd,
 reprint,
 amsmath,amssymb,
 aps,floatfix,
]{revtex4-2}

\usepackage{graphicx}
\usepackage{dcolumn}
\usepackage{bm}

\usepackage{quantikz}
\usepackage{pgfplots}
\newcommand{\I}{\mathrm{i}} 
\usepackage{amsmath, amssymb}
\usepackage{mathtools}
\usepackage{enumitem}  
\newcommand{\QED}{$\mathrm{QED}2$}
\newcommand{\QEDmat}{\mathrm{QED}2}

\RequirePackage{color}
\definecolor{dkgreen}{rgb}{0.2,0.7,0.4}
\definecolor{dkblue}{rgb}{0.2,0.2,0.7}
\definecolor{dkred}{rgb}{0.8,0,0}
\definecolor{dkgreen}{rgb}{0.2,0.8,0.4}

\usepackage{soul}

\begin{document}
\title{Adiabatic state preparation for digital quantum simulations of QED in $1+1$D}

\author{Matteo D'Anna}
\author{Marina Krsti\'{c} Marinkovi\'{c}}
\author{Joao C. Pinto Barros}
\affiliation{Institut f\"{u}r Theoretische Physik, ETH Z\"{u}rich,
Wolfgang-Pauli-Str. 27, 8093 Z\"{u}rich, Switzerland}

\date{\today}

\begin{abstract}
Quantum electrodynamics in $1+1$D (QED2) shares intriguing properties with QCD, including confinement, string breaking, and interesting phase diagram when the non-trivial topological $\theta$-term is considered. Its lattice regularization is a commonly used toy model for quantum simulations of gauge theories on near-term quantum devices. In this work, we address algorithms for adiabatic state preparation in digital quantum simulations of QED2.
We demonstrate that, for specific choices of parameters, the existing adiabatic procedure leads to level crossing between states of different charge sectors, preventing the correct preparation of the ground state. We further propose a new adiabatic Hamiltonian and verify its efficiency in targeting systems with a nonzero topological $\theta$-term and in studying string breaking phenomena.
\end{abstract}

\maketitle

\section{Introduction}

The study of strongly coupled quantum many-body systems poses unique challenges, often addressed through numerical simulations.
In particle physics, the archetypical example is provided by lattice Quantum Chromodynamics (lattice QCD), a gauge theory describing the strongly interacting sector of the Standard Model.
Due to the intrinsic non-perturbative nature of QCD, lattice regularization is the only \textit{ab-initio} approach to study this gauge theory at low energies. 

If we look beyond particle physics, gauge fields emerge from the collective behavior of interacting particles, e.g. in topological insulators or spin liquids~\cite{rokhsar1988superconductivity,moessner2001short,surace2020lattice,altland2010condensed,fradkin2013field}. 
Independently of their origin, the real-time dynamics of gauge theories, phase diagrams with non-zero topological $\theta$-term, and finite baryon density offer examples of sign or complex action problems. Their complexity has led to an intensive search for alternative methods, as they are generally intractable on classical computers using conventional simulation techniques for lattice gauge theories (LGTs) in large volumes (see e.g. \cite{troyer2005computational,Alexandru2022RevModPhys,nagata2022finite}). 

\par Quantum simulations offer a new paradigm to study gauge theories. This prospect 
motivates studies of many-body systems with a variety of gauge symmetry groups, representations, and dimensionalities to be built-up on different platforms (see e.g. \cite{banerjee2012atomic,zohar2013zold-atom,marcos2014two,zohar2015quantum,martinez2016real,bernien2017probing,muschik2017u,rico2018so,Gorg:2018xyc, schweizer2019floquet,pinto2020gauge,mil2020scalable,klco2020su2,banulsSimulatingLatticeGauge2020,davoudi2021search,Homeier:2020dvw,fontana2022reformulation,wang2022observation,fontana2023quantum,bauer2023quantum,kane2024nearlyoptimal,ciavarella2024quantum,fontana2024efficient,zache2023fermion, Cochran:2024rwe, gonzalez2024observation,de2024observation}). 
The abundance of studied models, often in low space-time dimensions,  provide a pathway towards quantum simulations of more intricate theories, like QCD~\cite{dimeglio2023quantum, Fauseweh:2024cow}.

In the case of digital quantum simulations, the degrees of freedom are mapped into qubits that can be manipulated to access static or dynamic properties. Regarding static properties, protocols like Adiabatic State Preparation (ASP) can be used to study areas of the phase diagram usually inaccessible through classical means \cite{chakrabortyClassicallyEmulatedDigital2022,hondaClassicallyEmulatedDigital2022,Ciavarella:2022skg}.
When it comes to dynamical properties, non-equilibrium physical regimes of interest can be studied, including, for example, thermalization \cite{surace2020lattice,dejongQuantumSimulationNonequilibrium2022} or the dynamical properties of string breaking \cite{surace2020lattice,zache2023fermion, Cochran:2024rwe, gonzalez2024observation,de2024observation}.

Finally, the development of quantum-based approaches for studying gauge theories has been accompanied by novel numerical techniques operating directly in the Hamiltonian formulation, including methods like quantum Monte Carlo \cite{frank2020emergence,banerjee2021recent,banerjee2024quantum,barros2024meron} and tensor networks \cite{banuls2020review,meurice2022tensor,montangero2022loop,kelman2024gauged,banuls2024parton,magnifico2024tensor}.

\par In this work, we address the preparation of the ground state in digital quantum simulations of the lattice Schwinger Model\:(\QED). With the ultimate goal of quantum simulations of the gauge theories of the Standard Model and beyond in mind, \QED~ is an important testing ground. It allows numerically cheaper and conceptually simpler studies of phenomena relevant for QCD, such as string breaking and phase diagram with a non-trivial topological $\theta$-term.
Driven by, among others, these motivations, there has been a recent surge of quantum simulations of \QED~\cite{dejongQuantumSimulationNonequilibrium2022,farrellQuantumSimulationsHadron2024,farrell202404,florioRealTimeNonperturbativeDynamics2023a,guo2024concurrentvqesimulatingexcited, nguyenDigitalQuantumSimulation2022, notarnicolaRealtimedynamicsQuantumSimulation2020, shawQuantumAlgorithmsSimulating2020}, 
including studies of non-zero topological angles \cite{angelidesFirstOrderPhaseTransition2023, banulsSimulatingLatticeGauge2020,borzenkovaVariationalSimulationSchwinger2021,chakrabortyClassicallyEmulatedDigital2022,funckeTopologicalVacuumStructure2020,Ghim:2024pxe,hondaClassicallyEmulatedDigital2022,hondaNegativeStringTension2022, ikedaDetectingCriticalPoint2023,Kaikov:2024acw,
 kharzeevRealtimeChiralDynamics2020, muellerQuantumComputationDynamical2023, pederivaQuantumStatePreparation2021, sakamotoEndtoendComplexitySimulating2023,schusterStudyingPhaseDiagram2023,4dd68de8a6ef3e975864603061a4f7522dbaef93,xieVariationalThermalQuantum2022,zacheDynamicalTopologicalTransitions2019} aimed at exploring phenomena such as topological transitions and charge screening. 
\QED~is quantum-simulated in experimental laboratories~\cite{SMexperimental1, SMexperimental2, SMexperimental3, SMexperimental4}
and classically,  using tensor networks~\cite{Buyens:2016ecr,buyensConfinementStringBreaking2016,Papaefstathiou:2024zsu}
and Euclidean LGT approaches~\cite{Gattringer:2015nea}.
\par We build on the existing algorithms for quantum simulations of \QED~and introduce a new method for adiabatic state preparation of its ground state. 
We argue that conventional approaches to ASP~\cite{chakrabortyClassicallyEmulatedDigital2022, hondaClassicallyEmulatedDigital2022, pederivaQuantumStatePreparation2021,Ghim:2024pxe} are not suited for a generic ground state preparation of \QED~with non-zero $\theta$ angle. In \QED, with open boundary conditions (which we use since they allow to completely integrate out of gauge fields \cite{hamer1997series}) and for a particular choice of the Hamiltonian's parameters, the lowest energy state will have a non-zero charge. 
The usually applied ASP approaches preserve the charge of the initial state of adiabatic evolution, and thus become problematic once it is not in the charge sector of the true ground state. Even worse, it is also possible that the mentioned ASP approaches fail, due to some level crossing, to prepare the correct ground state even in the case that the initial state was in the correct charge sector.
We propose a new adiabatic Hamiltonian, which successfully avoids the issues of level crossing and ground state preparation for arbitrary charge sectors. The new ASP procedure therefore allows us to explore the phase diagram and string breaking for non-zero topological $\theta$ angles, in both cases that we restrict to a given charge sector or we look for the ground state of the whole Hilbert space.

\par The rest of the text is organized as follows: in Section \ref{sec:SM} we review the formulation of the Schwinger Model with $\theta$-term, including the derivation of a lattice Hamiltonian and the setup used to study the string breaking dynamics. 
Section \ref{sec:ASP} introduces adiabatic state preparation algorithms. Here, we review the existing procedure that keeps the lattice charge operator fixed and discuss its limitations. We then introduce a new adiabatic Hamiltonian based on the symmetries of the system, which mixes states from different charge sectors and thus allows a reliable ground state estimation for a wide parameter set of topological $\theta$-angles and masses. 
In Section \ref{sec:res} we test the proposed method numerically and apply it to
study the string breaking dynamics in \QED, with final remarks and outlook provided in Section \ref{sec:outro}. 

\section{The Model}\label{sec:SM}
\subsection{The lattice Hamiltonian}
We consider a lattice Hamiltonian of the Schwinger Model \cite{schwingerGaugeInvarianceMass1962, colemanChargeShieldingQuark1975, colemanMoreMassiveSchwinger1976}, which is obtained from the Lagrangian after a chiral rotation \cite{chakrabortyClassicallyEmulatedDigital2022}

\begin{equation}
    \label{eq:Hqedlat}
    \begin{aligned}
H= & -\I \sum_{n=1}^{N-1}\left(w-(-1)^n \frac{m}{2} \sin \theta\right)\\ &\left(\chi_n^{\dagger} e^{\I \varphi_n} \chi_{n+1}-\chi_{n+1}^{\dagger} e^{-\I \varphi_n} \chi_n\right) \\
& +m \cos \theta \sum_{n=1}^N(-1)^n \chi_n^{\dagger} \chi_n+J \sum_{n=0}^{N-1} L_n^2.
\end{aligned}
\end{equation}
The bare parameters of this model are the gauge coupling $g$, the fermion mass $m$, and the topological angle $\theta$. Additionally, $w= 1/(2a)$ and $J=g^2a/2$ depend on the lattice spacing $a$.
The lattice Hamiltonian \eqref{eq:Hqedlat} commutes with a set of local operators,
\begin{equation}
G_n=\left(L_n-L_{n-1}\right)-\left(\chi_n^{\dagger} \chi_n-\frac{1-(-1)^n}{2}\right),
\end{equation}
which also commute with each other.
As a consequence, the Hilbert space can be broken down into different sectors where states are classified according to a Gauss's law $G_n\left|\psi\right>=\rho_n\left|\psi\right>$. 
Here we will be interested in the cases where $\rho_n=0\ \forall n$, or when two static charges (with total charge zero) are present at two different sites, $n_+$ and $n_-$, i.e. $\rho_n=q\delta_{nn_+}-q\delta_{nn_-}$. We will refer to these sectors as \emph{physical sectors}. In the latter expression, $n_+$ ($n_-$) is the position of the positive (negative) charge and $q\in\mathbb Z$. We will refer to $m, g, \theta, a, n_\pm,q$ as the parameters of our theory (\QED~parameters), where instead of the lattice spacing $a$, we will occasionally use the lattice extent $L = a(N-1)$. 
Throughout the text, we set $q=1$.

We will consider solely open boundary conditions, as this choice allows to integrate out 
the gauge field degrees of freedom by an iterative application of Gauss's law,
\begin{equation}\label{eq:GaussLaw}
    G_n=q \delta_{n n_{+}}-q \delta_{n n_{-}},
\end{equation}
and results in
\begin{equation}
\label{eq:GaussSol}
    \begin{split}
    L_n=&L_0+\sum_{k=1}^{n}\left(\chi_{k}^{\dagger} \chi_{k}-\frac{1-(-1)^{k}}{2}\right)\\
    & +q \Theta\left(n-n_{+}\right)-q \Theta\left(n-n_{-}\right).
    \end{split}
\end{equation}
$\Theta\left(n-n_{+}\right)$ denotes the Heaviside step function with the condition $\Theta\left(0\right)=1$. The model with $(\theta, L_0)$ is equivalent to the one with $(\theta+2\pi L_0, 0)$ \cite{colemanMoreMassiveSchwinger1976} and we interpret the $\theta$-term as a background field. Therefore, without loss of generality, we always set the first link to $L_0=0$.

\par The angles $\varphi_k$ can be absorbed in the fermionic variables through
\begin{equation}\label{eq:absorbphi}
    \chi_{n} \rightarrow\left(\prod_{l<n} e^{-\I \varphi_{l}}\right) \chi_{n},
\end{equation}
leaving \eqref{eq:Hqedlat} as a pure fermionic system with long-range interactions. Such system can be further recast as a spin system through the Jordan-Wigner transformation \begin{equation}\label{eq:JW}\chi_n=\left(\prod_{l<n}-i Z_l\right) \frac{X_n-i Y_n}{2},\end{equation}which is well suited for the quantum computing framework~\cite{chakrabortyClassicallyEmulatedDigital2022}.
The \QED~Hamiltonian takes the form
\begin{equation}
    \label{eq:HSM}
    H_\mathrm{\QEDmat}= H_{ZZ} + H_\pm + H_Z,
\end{equation}
with
\begin{equation}
    \begin{aligned}
H_{ZZ}  = &\frac{J}{2} \sum_{n=2}^{N-1} \sum_{1 \leq k<\ell \leq n} Z_k Z_{\ell}, \\
H_{ \pm}  =&\frac{1}{2} \sum_{n=1}^{N-1}\left(w-(-1)^n \frac{m}{2} \sin \theta\right) \\ &\left(X_n X_{n+1}+Y_n Y_{n+1}\right), \\
H_Z  =&\frac{m \cos \theta}{2} \sum_{n=1}^N(-1)^n Z_n\\ &+J \sum_{n=1}^{N-1}v(n; n_+,n_-)\sum_{\ell=1}^n Z_{\ell},
\end{aligned}
\end{equation}
where
\begin{equation}\label{eq:vorf}
    \begin{split}
    v(n; n_+,n_-)= &-\frac{n \bmod 2}{2} +q\Theta(n-n_+)\\ &-q\Theta(n-n_-)
    \end{split}
\end{equation}
is the only factor affected by the presence and position of the external static charges. The all-to-all interaction expressed in $H_{ZZ}$ results from solving Gauss's law using~\eqref{eq:GaussSol}.
We refer to \cite{chakrabortyClassicallyEmulatedDigital2022} and references therein for a detailed derivation of the lattice Hamiltonian \eqref{eq:HSM} starting from the \QED~Lagrangian in the continuum.

\subsection{Charge sectors}
\par The charge operator on the lattice is given by
\begin{equation}\label{eq:Qtot}
    Q = \frac g2 \sum_{n=1}^N Z_n,
\end{equation}
and commutes with the Hamiltonian for any choice of \QED~parameters. We can classify eigenstates of $H_{\QEDmat}$ into different charge sectors. The charge eigenvalues are $\{-Ng/2, {-(N/2-1)g},\cdots, Ng/2\}$, which characterize the different charge sectors. We will always refer to charge sectors in units of $g$.

\par With periodic boundary conditions, all physical states are in the charge $0$ sector. To be able to integrate out the gauge fields, we use open boundary conditions in this work. This no longer restricts physical states, including the ground state, to be in the zero charge sector.
It is common, for the region $1\lesssim\theta/\pi\lesssim3/2$, to have the ground state in the charge $1$ sector.
Additionally, the presence (and position) of the static charges can influence the charge sector of the ground state as well.
We can therefore be interested in two situations: investigate the ground state of the whole Hilbert space (that is, the ground state across charge sectors), or restrict it to a specific charge sector. For the latter, one can add to the \QED~Hamiltonian the term 
\begin{equation}\label{eq:Qshift}
\begin{split}
    H_\text{shift} &= \lambda_Q\left(\frac Qg -Q_t\right)^2\\
    &= \lambda_Q\left(-Q_t\sum_{n=1}^NZ_n+\frac12\sum_{n=2}^N\sum_{l=1}^n Z_nZ_l\right),
    \end{split}
\end{equation}
where factors proportional to the identity were omitted in the second line. The parameter $Q_t\in \{-N/2, -N/2+1, \cdots, N/2\}$ is an integer indicating the desired charge sector and $\lambda_Q\in \mathbb R_{\geq 0}$ is an appropriate scaling factor. It should be chosen large enough such that the ground state of $H_{\QEDmat}+H_\text{shift}$ lies in the $Q_t$ charge sector. 

\subsection{String breaking}

\par Ground state preparation allows us to study string breaking, a key feature of gauge theories exhibiting confinement.
This is particularly relevant when there is a non-zero topological $\theta$ angle and standard classical approaches suffer from a sign problem. 
We study the ground state of the \QED~Hamiltonian \eqref{eq:HSM} for varying positions $n_\pm$ of the static charges. For small $\theta$, when the distance $d = a\left|n_+-n_-\right|$ is small, increasing it will lead to a cost of energy. If this energy becomes large enough, one expects that pair-creation is favorable. This will screen the static charges leading to a plateau of the energy if the distance keeps being increased, signaling string breaking.

\par While restrictions on current emulators and near-term quantum hardware place us in a situation where $d$ can never be too large, and boundary effects are significant, these sizes are enough to observe qualitative effects of string breaking.
Following \cite{buyensConfinementStringBreaking2016} we keep track of two quantities. The first one is the potential
\begin{equation}
    V(d) := E_0(d)-E_0(0),
\end{equation}
where $E_0(d)$ is the ground state energy of the \QED~Hamiltonian when the static charges are at a distance $d$. The second one is the charge on the first half of the chain $Q_-$, given by
\begin{equation}\label{eq:Qmin}
    Q_- = \frac g2 \sum_{n=1}^{N/2} Z_n.
\end{equation}
At $\theta=0$ the expected behavior for $V(d)$ corresponds to linear growth until pair creation becomes favorable, plateauing afterward, as described above. Since $V(d)$ is strongly influenced by boundary effects  (following \cite{hondaClassicallyEmulatedDigital2022} the plateau occurs in a region where $1\ll gd \ll gL$), we rely on $Q_-$ in order to look for evidence of string breaking. If we place the two external static charges in the middle of the chain and then move them towards the respective boundaries, a pair-creation will be registered by $Q_-$ as a jump in its value.

\section{Adiabatic state preparation}\label{sec:ASP}
We start this Section by briefly reviewing the adiabatic condition. Afterward, we examine the standard ASP procedure and elucidate why, for certain choices of \QED~parameters, it fails to produce the lowest energy state of the charge $0$ sector. To conclude the Section, we introduce a new adiabatic Hamiltonian more suited to prepare both the true ground state and the lowest energy state within a target charge sector.

\subsection{The adiabatic condition}

The adiabatic theorem ensures that if we prepare our system in an eigenstate of the Hamiltonian and vary the latter slowly enough, the state will evolve in an eigenstate at all times, as long as a gap to other states is preserved.

Concretely, let us consider a time-dependent Hamiltonian $H_A\left(t\right)$ with eigenstates $\ket{k_t}$. We further assume the existence of a single ground state labeled by $k=0$, prepared at $t=0$. It is posteriorly left to evolve, under $H_A(t)$, until $t=T$. The adiabatic condition is fulfilled as long as \cite{albashAdiabaticQuantumComputation2018} 
\begin{equation}\label{eq:adiabcond}
\frac{1}{T} \max _{s \in[0,1]} \frac{\left|\left\langle 0_{s\cdot T}\left|\partial_s \widetilde H_A(s)\right| k_{s\cdot T}\right\rangle\right|}{\Delta(s)^2} \ll 1, \quad\forall k \neq 0,
\end{equation}
where it is assumed that it is possible to write the Hamiltonian $H_A$ in a ``timescale-independent'' way, such that $\widetilde H_A(s) = H_A(sT)$ with $s\in [0,1]$ does not depend on $T$.
We denote by ${\Delta (s):= E_{1_s}-E_{0_s}}$ the spectral gap.
\par If the adiabatic condition \eqref{eq:adiabcond} is satisfied, the ground state at $T$ is approximated by 
\begin{equation}\label{eq:SMgs}
    \ket{0_T}\approx \mathcal{T} \exp \left(-\mathrm{i} \int_0^T \mathrm d t H_A(t)\right)\ket{0_0},
\end{equation}
with equality being given in the limit ${T\to\infty}$. We approximate the time-ordered exponential by
\begin{equation}\label{eq:SMgscirc}
    \ket{0_T} \approx \ket{\varphi_T}:= \underbrace{U(T) U(T-\delta t) \cdots U(2 \delta t) U(\delta t)}_{M\text{ steps}} \ket{0_0},
\end{equation}
where $U(\tau) := \exp(-\mathrm i H_A(\tau)\delta t)$ and $\delta t=T/M$. We refer to $T, M, \delta t$ as the adiabatic parameters.

\par If the gap is small or closing, the adiabatic condition is not satisfied. If the gap is small but non-vanishing, we typically experience some leakage into the first excited states, and the prepared state is a superposition of multiple eigenstates. The mixing does not occur if a symmetry of the adiabatic Hamiltonian protects the states involved.
However, in the latter case, the tracked state might not be the true ground state.

\subsection{ASP restricted to the 0 charge sector}
\par An ASP procedure that conserves the charge of the initial state is given by the Hamiltonian proposed by \cite{chakrabortyClassicallyEmulatedDigital2022}. It corresponds to \eqref{eq:HSM} with the substitutions
\begin{equation}\label{eq:HA1subs}
    \begin{split}
        m & \to m(t) :=m_0\left(1-\frac tT\right) + m\frac tT \\
        w & \to w(t):=w\frac tT\\
        \theta & \to \theta(t):=\theta \frac tT,
    \end{split}
\end{equation}
for $m_0\in \mathbb R$.
Equivalently, this corresponds to
\begin{equation}\label{eq:HA1}
    H_{A1}(t) = H_{\QEDmat}\big\vert_{m=m(t), w=w(t), \theta = \theta(t)}.
\end{equation}
Because $[H_{A1}(t),Q]=0\ \forall t$, the adiabatic evolution will conserve the total charge.

In the absence of static charges the ground state of $H_{A1}(0)$ is $(\ket 0 \ket 1)^{\otimes N/2}$, which is a state with total charge $0$ \cite{chakrabortyClassicallyEmulatedDigital2022}.
This adiabatic evolution is not well suited for preparing the ground state of \QED, if the true ground state will not remain in the charge 0 sector. For $\theta\neq 0$, and depending on boundary conditions, this is not always the case.
The presence of static charges can potentially worsen the situation: first of all, the adiabatic Hamiltonian is modified in a way that the ground state for $t=0$ can no longer be simply determined. Concretely the term $v(n; n_+,n_-)\sum_{\ell=1}^n Z_{\ell}$ in
\begin{equation}\label{eq:HA0full}
    \begin{split}
    H_{A1}(0) =& \frac{m_0\cos(\theta_0)}{2} \sum_{n=1}^N(-1)^n Z_n\\ &+J \sum_{n=1}^{N-1}v(n; n_+,n_-)\sum_{\ell=1}^n Z_{\ell} \\
    & +\frac{J}{2} \sum_{n=2}^{N-1} \sum_{1 \leq k<\ell \leq n} Z_k Z_{\ell},
    \end{split}
\end{equation}
can become prominent.
In the worst case determining the ground state of \eqref{eq:HA0full} requires to check the $2^N$ states forming the computational basis.
Additionally, when string breaking happens, we typically observe that the ground state of \QED~changes charge sector.

In our approach, we will prepare an initial state that is in a superposition of all sectors of interest, addressing directly this first obvious obstacle of the original approach outlined above.

\subsection{ASP for states within arbitrary charge sectors}
In order to address the issues raised above we introduce the new adiabatic Hamiltonian
\begin{equation}\label{eq:HA2}
H_{A2}(t) = \left(1 -\frac tT\right)\beta\sum_{n=1}^N (-1)^n X_n + \frac tT H_{\QEDmat},
\end{equation}
where $\beta\in \mathbb{R}_{>0}$ is a tunable parameter. It allows to prepare states in different charge sectors, and therefore can be used for different choices of \QED~parameters. The new adiabatic path is of the form
\begin{equation}
H_{A}(t) = g(t) H_0 + h(t) H_\mathrm{target},
\end{equation}
for $g, h$ such that $g(0)=h(T)=1, g(T)=h(0)=0$.
The functions $g,h$ can be explored in order to improve the results and/or accelerate the convergence.
Here we only consider the linear case, $h\left(t\right)=t/T$ and $g\left(t\right)=1-h\left(t\right)$.
The initial Hamiltonian is given by
\begin{equation}\label{eq:H0}
H_0=\beta \sum_{n=1}^N (-1)^n X_n,
\end{equation}
which has ground state $(\ket + \ket -)^{\otimes N/2}$, independently of any choice of \QED~parameters.
The adiabatic evolution given by \eqref{eq:HA2} does not preserve the charge for $t\neq T$
\begin{equation}
 [H_{A2}(t), Q] = (-2\mathrm{i}) \left(1-\frac tT\right)\beta \frac g2\sum_{n=1}^N  (-1)^nY_n,   
\end{equation}
and therefore allows, in principle, for the preparation of states in different charge sectors. We will observe that \eqref{eq:HA2} can produce the correct ground state for all choices of parameters checked.
A rigorous assertion regarding the effectiveness of $H_{A2}$ requires proving that for any $s\in[0,1)$ the adiabatic Hamiltonian \eqref{eq:HA2} has a non-zero gap, thus making it possible to choose a finite $T$ to satisfy the adiabatic condition \eqref{eq:adiabcond}.
This is typically a challenging task, which is beyond the scope of this paper.

\section{Results}\label{sec:res}
In this Section we compare the vacuum expectation values (VEVs) obtained with the two adiabatic procedures, and observe that, for arbitrary choices of \QED~parameters, only the newly introduced adiabatic Hamiltonian correctly reproduces the results of exact diagonalization.

\subsection{Setup}
In order to carry out the ASP procedure \eqref{eq:SMgscirc} we first need to choose a Suzuki-Trotter decomposition for $U_{j=1,2}(\tau)$, where $U_j$ is the unitary evolution given by the adiabatic Hamiltonian $H_{Aj}$. We choose the first order Suzuki-Trotter decomposition \cite{jonesOptimisingTrottersuzukiDecompositions2019}, and $U_{1,2}$ are given by 
\begin{equation}\label{eq:U12}
    \begin{split}
        U_1(\tau)  =& \left(e^{-\mathrm i (H_{ZZ}(\tau)+H_Z(\tau))\delta t}e^{-\mathrm i H_{YY}(\tau)\delta t}\right. \\
         & \left.e^{-\mathrm i H_{XX}(\tau)\delta t}\right)+ \mathcal O\left(\delta t^2\right) \\
        U_2(\tau)  =& \left(e^{-\mathrm i (H_{ZZ}+H_Z)\delta t\cdot \tau/T}e^{-\mathrm i H_{YY}\delta t\cdot\tau/T}\right.\\ &\left.e^{-\mathrm i (H_{XX}\delta t\cdot \tau/T +H_0\delta t\cdot (1-\tau/T))}\right) + \mathcal O\left(\delta t^2\right), \\
    \end{split}
\end{equation}
where each exponential contains Pauli operators that are all commuting with each other.
A complete description of the gates needed for producing the terms appearing in the \QED~Hamiltonian is given in \cite{chakrabortyClassicallyEmulatedDigital2022}. For implementing $U_2$ we additionally need a decomposition for $\exp(-\mathrm i H_0 t)$
\begin{equation}
\begin{split}
    e^{-\mathrm i H_0 t} &= \prod_{n=1}^N e^{-\mathrm i(-1)^{n}\beta t X_n}\\
     &= \prod_{n=1}^N \mathrm{RX}_n\left((-1)^n2\beta t\right),
     \end{split}
\end{equation}
where $\mathrm{RX}_n$ indicates the $\mathrm{RX}$ gate acting on qubit $n$.

\par The final state $\ket{\varphi_{j, T}}\approx \ket{0_T}$, resulting from the application of $M$ steps of $U_j(\tau)$, is prepared on a register of $N$ qubits following \eqref{eq:SMgscirc}. We compute the estimate of the VEV $\langle O\rangle_{\varphi_{j, T}}$ for different observables $O$. The results for the ground state energy and the total charge \eqref{eq:Qtot} are discussed in main text.
\par  The simulations are carried out using IBM's Qiskit Python library \cite{Qiskit}. We compare the simulation results with exact diagonalization, obtained with the Python library QuSpin \cite{Quspin}.

As mentioned above, $\beta$ is the only tunable parameter of the initial Hamiltonian \eqref{eq:H0}. The efficacy of the approach will depend on this choice, but it does not require fine-tuning. One would expect that the best choice for $\beta$ would lie in an interval where the initial and final ground states have similar energy. We find this to be the case only when very expensive Suzuki-Trotter decompositions are employed.
The systematic study of the optimal choice of $\beta$ for arbitrary system sizes is beyond the scope of this paper, but for the volumes studied here, we observe it to lie in the interval $\beta N\in [20, 40]$.

\subsection{Numerical simulations}
\par
\begin{figure*}
\includegraphics[width=\textwidth]{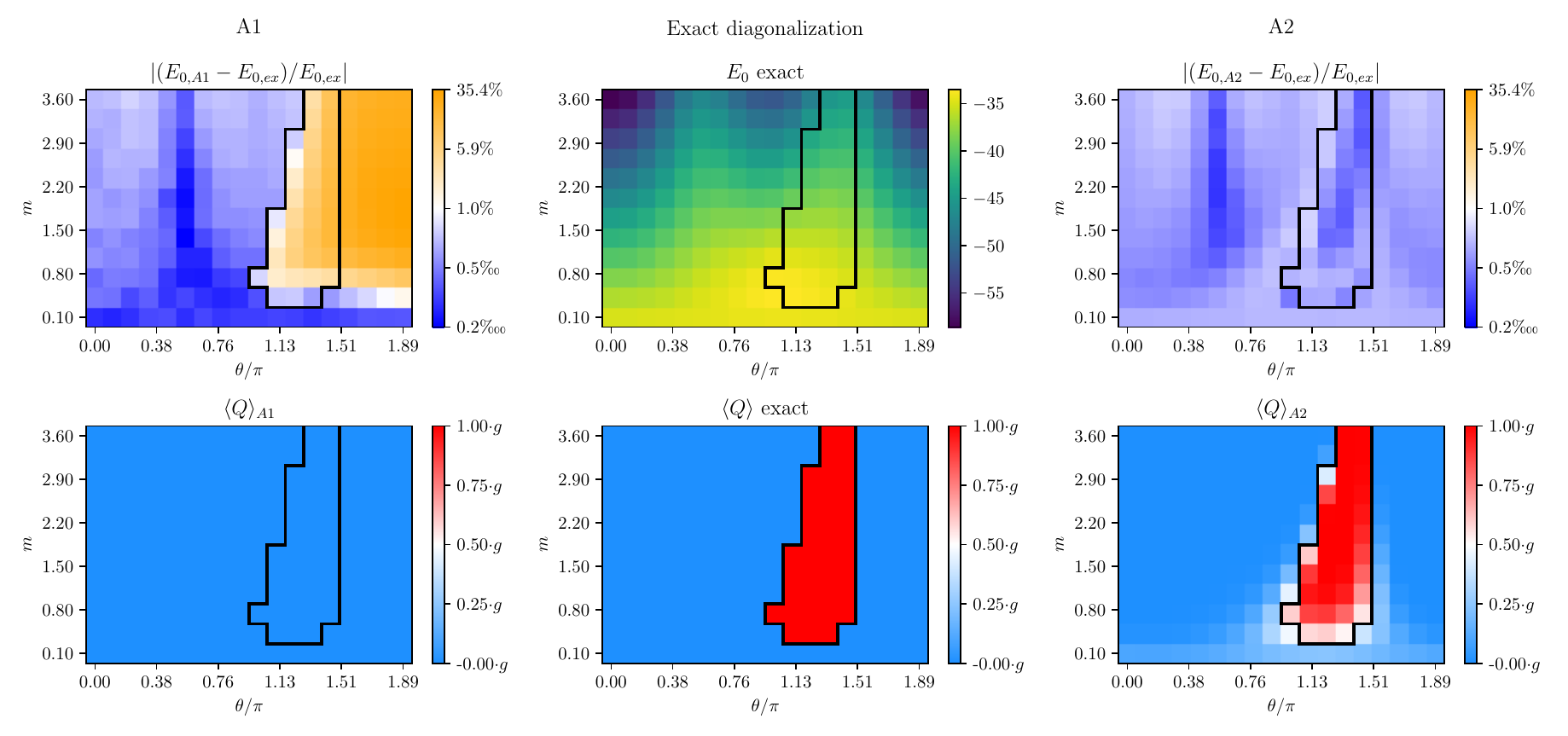}
\caption{Expectation values of $H_{\QEDmat}$ (first row) and $Q$ (second row) for a state prepared with $H_{A1}$ (left column), $H_{A2}$ (right column) for different choices of $m,\theta$. This can be compared with the result of exact diagonalization (center column). We consider a lattice with $16$ sites, $L=8.19, g=1.8$, in the absence of static charges. For $H_{A2}$ we set $\beta N=30$. For $H_{A1}$ we choose $m_0=0.5, \theta_0=0$.
For both adiabatic procedures we used $M=1000$ steps and $T=60$.}
\label{fig:allcomp}
\end{figure*}

 We first compute the ground state energy and the charge VEV with $H_{Aj=1,2}$ for different choices of $m,\theta$, and we compare them with the results from exact diagonalization. The results for a lattice with $N=16$ sites are shown in Figure \ref{fig:allcomp}.

As expected from the discussion above, the state prepared with $H_{A1}(t)$ never leaves the $0$ charge sector. 
This does not correspond to the ground state for a large region of the $m$-$\theta$ parameter space, where the ground state is in the charge $1$ sector. For $H_{A1}$ we also notice that the ground state energy estimation is the most inaccurate in the region where $\theta > 1.5\pi$, even though the ground state is in the 0 charge sector. This is the result of the adiabatic path starting at $\theta(0)=0$ and crossing the charge 1 region. Hence it is possible to prepare the correct ground state for the $\theta > 1.5\pi$ region starting with $\theta(0)=2\pi$, but ground state preparation in the charge 1 region remains inaccessible for $H_{A1}$.
In turn, $H_{A2}$ can reproduce the correct energy across different charge sectors, with a relative error at worst of order $10^{-2}$.

\par
\begin{figure*}
\includegraphics[width=.995\textwidth]{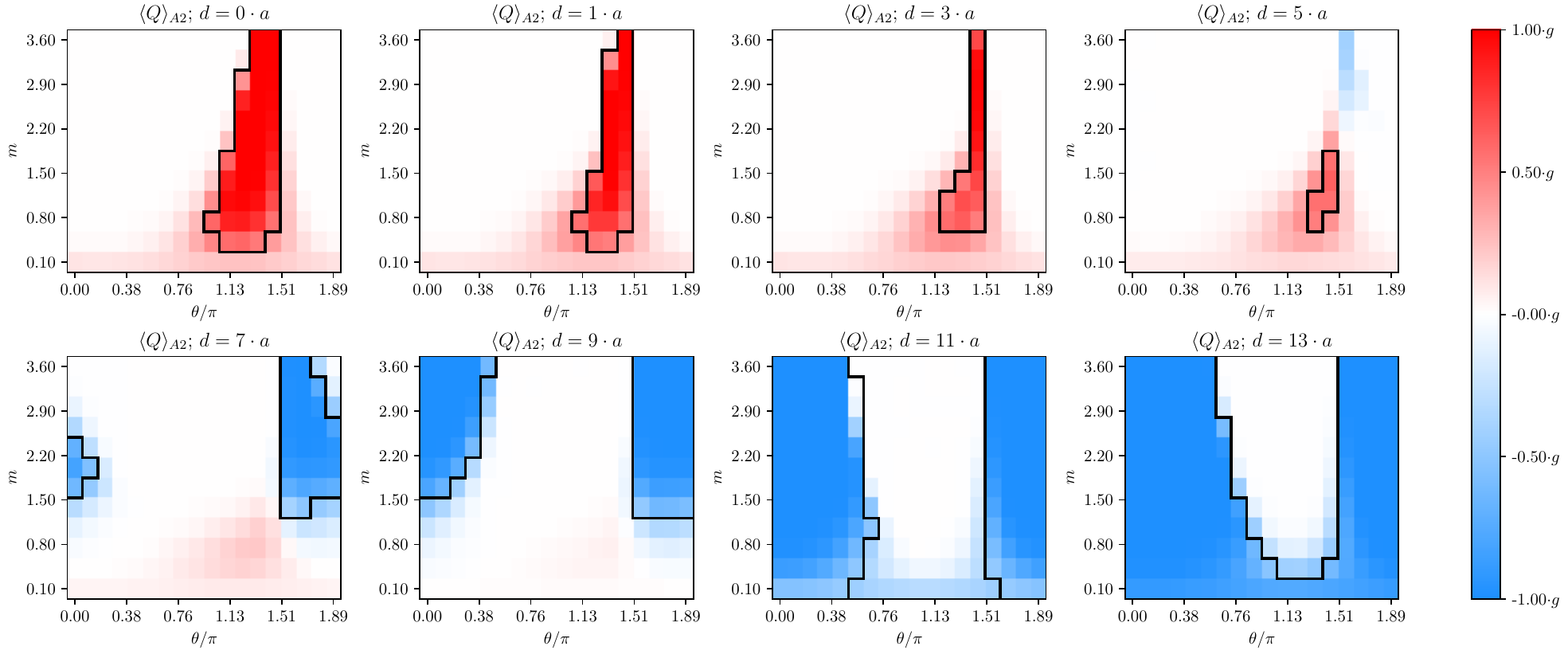}
\caption{Comparison of the charge VEV obtained using $H_{A2}$ with exact diagonalization for a lattice with $16$ sites, and for various positions of the static charges. All other \QED~and adiabatic parameters are identical to the setup described in Figure \ref{fig:allcomp} ($L=8.19,~g=1.8,~\beta N=30,~M=1000,~T=60$). The black lines refer to the boundaries of the charge regions obtained from exact diagonalization.}
\label{fig:Qcomp}
\end{figure*}

 We also study the effect of introducing static charges. We start by placing them in the middle of the lattice, $n_+=N/2, n_-=N/2+1$, and increase their distance by moving each of them one lattice site at a time.
The charge VEVs obtained with $H_{A2}$, together with exact diagonalization results, are presented in Figure \ref{fig:Qcomp}.
The adiabatic procedure $H_{A2}$ gives results that are in agreement with exact diagonalization for most setups of the static charges that were explored, and it does so independently of the charge of the two regions. 
The charge VEV from $H_{A2}$ is only inaccurate around the boundaries of the two charge regions. This is because there the gap between the ground states of the two charge sectors is closing (and vanishes at the boundary).
As a consequence, there will be some leakage to the other sector.
It is interesting to note that this leakage can be detected within this procedure, since it will result in a non-integer charge expectation value.
To improve the accuracy of the result for that specific choice of parameters the ASP procedure should be repeated by choosing a larger $T$.
The charge expectation value can thus be used as a probe for the success of the state preparation procedure.
\begin{figure}
\includegraphics[width=.48\textwidth]{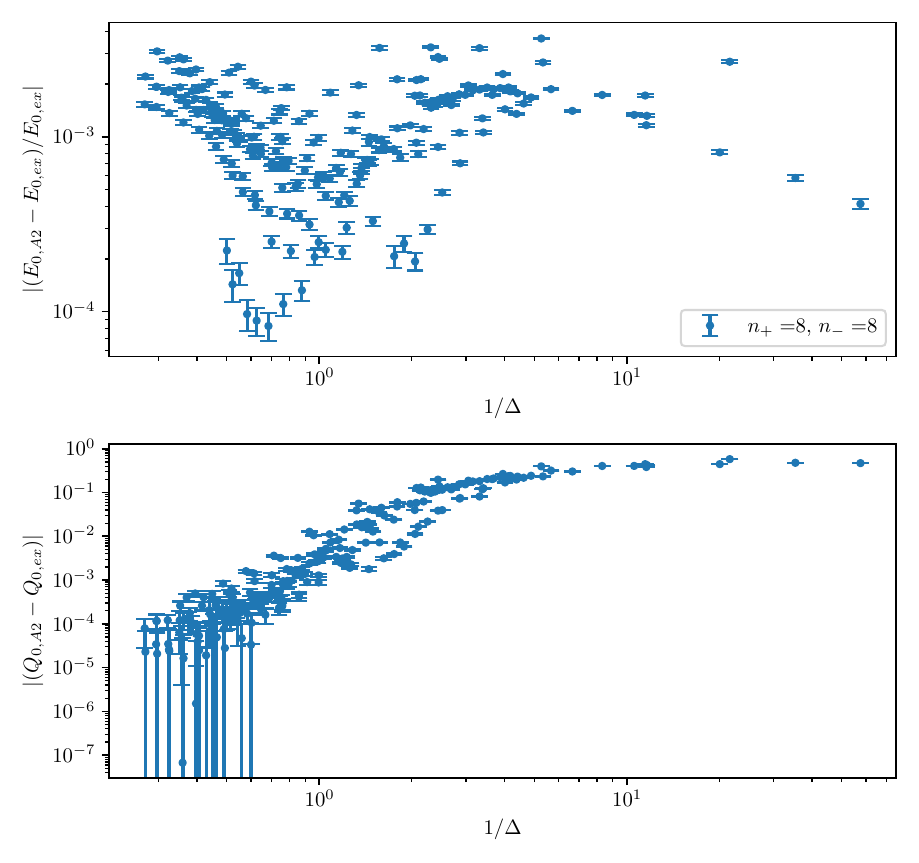}
\caption{Top: relative error of ground state energy estimation as a function of $1/\Delta$, where $\Delta$ is the gap of the \QED~Hamiltonian obtained from exact diagonalization. Bottom: difference between estimated and exact charge vev as a function of $1/\Delta$. All parameters are identical as in Figure \ref{fig:allcomp} ($N=16,~L=8.19,~g=1.8,~\beta N=30,~M=1000,~T=60$).}
\label{fig:gapcorr}
\end{figure}
This is illustrated in Figure \ref{fig:gapcorr}, where we plot the relative error in the ground state energy and the error in the charge as a function of the gap.
Figure \ref{fig:gapcorr} indicates that large errors in the estimations in the charge strongly correlate with smaller gaps, but strong correlations are not observable between the gap and relative error of the ground state energy.
Repeating the simulations including static charges leads to similar behavior: larger gaps correlate with larger errors in the estimation of the charge VEV, but do not correlate with larger errors in the ground state energy. We emphasize that this observation is possible with $H_{A2}$, while the same analysis cannot be performed for states prepared with $H_{A1}$, as leakage to other sectors is by definition prohibited for $H_{A1}$.

\par
\begin{figure*}
\includegraphics[width=.95\textwidth]{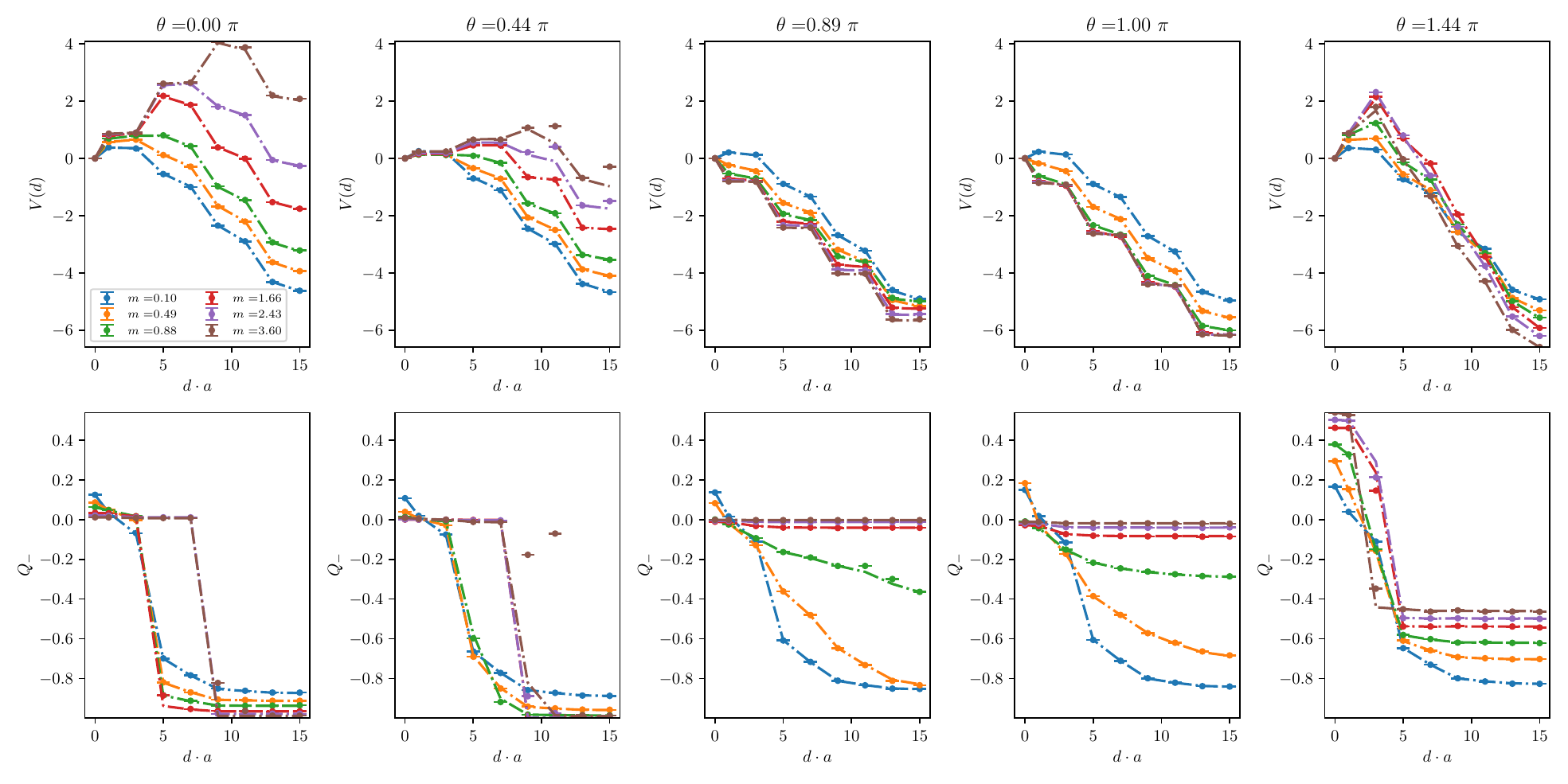}
\caption{VEVs of potential $V(d)$ (top row) and  charge $Q_-$ (bottom row) for the lowest energy state of the charge $0$ sector. The \QED~parameters are $N=16$, $ L=8.19$, $g=1.8$, the ``shift parameter'' is $\lambda_Q=4$, and the adiabatic parameters are $M=1000$, $T=60$, $\beta N=30$. The lines refer to the result from exact diagonalization, and the dots are obtained by the ASP procedure with $H_{A2}$.}
\label{fig:SB}
\end{figure*}

 States prepared with $H_{A2}$ can also be restricted to a given charge sector. We illustrate this by studying string braking in the presence of static charges. To do this, we consider the \textit{shifted} \QED~Hamiltonian $H_{\QEDmat}+H_\text{shift}$ (see Eq. \eqref{eq:Qshift}). The plots of $V$ and $Q_-$, as a function of $d$, for different $\theta,m$, are in Figure \ref{fig:SB}. As discussed in Section \ref{sec:SM}, boundary effects prevent us from obtaining the typical shape of the potential that we would expect from string breaking.
On the other hand, $Q_-$ can be used to detect the distance at which string breaking happens. Once it occurs, we expect that charges are accumulated in half of the lattice, even if the total charge of the system is zero. This is observed for all topological angles that we have probed, as long as the mass is small enough. For angles close to $0$ (or $2\pi$), this change is abrupt providing a strong signal of string breaking.
The results are in agreement with exact diagonalization. We only get inaccurate results for the points with $\theta=0.44\pi, m=3.60$ and $d/a= 9, 11$, because, for $\lambda_Q=0$ we are very close to the boundary of the two charge regions, hence the gap is smaller than usual. This issue can be resolved by choosing a larger $\lambda_Q$ and/or a larger $T$.

\section{Conclusions and Outlook}\label{sec:outro}
\par Adiabatic state preparation allows for the preparation of ground states of strongly coupled systems. One of the key factors for its success is the judicious choice of the adiabatic path. If the system possesses a symmetry, its spectrum is broken down into different symmetry sectors. If this symmetry is present throughout the entire adiabatic path, we are no longer guaranteed to find the true ground state of the system, as level crossing can occur across sectors. Preparing ground states of gauge theories is bound to be an important task in understanding these models beyond classical limitations. Such protocols will enable the computation of static and dynamic properties, where standard Monte Carlo encounters a sign problem. The regimes studied in these papers are relevant in both cases. Our protocol allows for the preparation of the ground states for a generic $\theta$-term, which is a source of a sign problem. This can be done at the same time that we add static charges, enabling the computation of other static quantities like the string tension, addressing the presence of confinement. Furthermore, it also allows for the preparation of states to undergo a quench protocol. In that case, we can have access to dynamical properties of string breaking, which is another paradigmatic phenomenon of confining theories. Complementary to recent notable experimental realizations \cite{gonzalez2024observation,de2024observation}, our results enable the study of string-breaking quenching from the system's ground state across different $\theta$ angles and within different charge sectors.   

\par  Concretely, we propose to use a new adiabatic Hamiltonian for preparing the ground state of the \QED~Hamiltonian. By not conserving charge until the very end of the adiabatic path, we can prepare states with different charges. This is crucial to allow for the preparation of the ground state in the entire region of \QED~parameter space. We validate the new approach by comparing the VEVs of observables of interest with both the standard adiabatic Hamiltonian present in the literature and the results from exact diagonalization. We find that unlike the standard procedure for ASP, the version proposed in this work can reproduce the reference results for all choices of \QED~parameters. This is particularly relevant since a standard adiabatic path can fail to reproduce correct results at a finite $\theta$ angle. We also investigate \QED~with external static charges. The new ASP procedure performs well also in this case without any adjustment needed when compared to the absence of static charges.
 
\par The fact that the ground state can be found in different charge sectors in this work is a consequence of the chosen boundary conditions, which allow for the integration of gauge fields. The new protocol for ASP is then especially relevant in the case of open boundary conditions. Another interesting application pertains to the use of $C^*$ boundary conditions. Like open boundaries, these boundary conditions allow for charged states but, in contrast to the open boundaries, the $C^*$ approach retains translation invariance. For this reason, $C^*$-boundary conditions are being used in state-of-the-art simulations of QCD+QED~\cite{RCstar:2022yjz}, making it interesting to test our protocol against such a scenario in \QED. Finally, an important step towards applying the proposed ASP protocol in regimes with fewer classical methods available is the extension to higher dimensions and other gauge groups.

{\it Acknowledgements.}---We are grateful to Debasish Banerjee, Christian Bauer, Dorota Grabowska, and Yannick Meurice for insightful discussions. This research was supported by the Munich Institute for Astro-, Particle and BioPhysics (MIAPbP), which is funded by the Deutsche Forschungsgemeinschaft (DFG, German Research Foundation) under Germany's Excellence Strategy – EXC-2094 – 390783311. We acknowledge access to Piz Daint at the Swiss National Supercomputing Centre, Switzerland under the ETHZ's share with the project ID eth8. Support from the Google Research Scholar Award in Quantum Computing and the Quantum Center at ETH Zurich is gratefully acknowledged. We acknowledge use of the IBM Q for this work. The views expressed are those of the authors and do not reflect the official policy or position of IBM or the IBM Q team.

\appendix
\setcounter{figure}{0}
\renewcommand\thefigure{S\arabic{figure}}

\section{Convergence studies}
\begin{figure*}
\includegraphics[width=.95\textwidth]{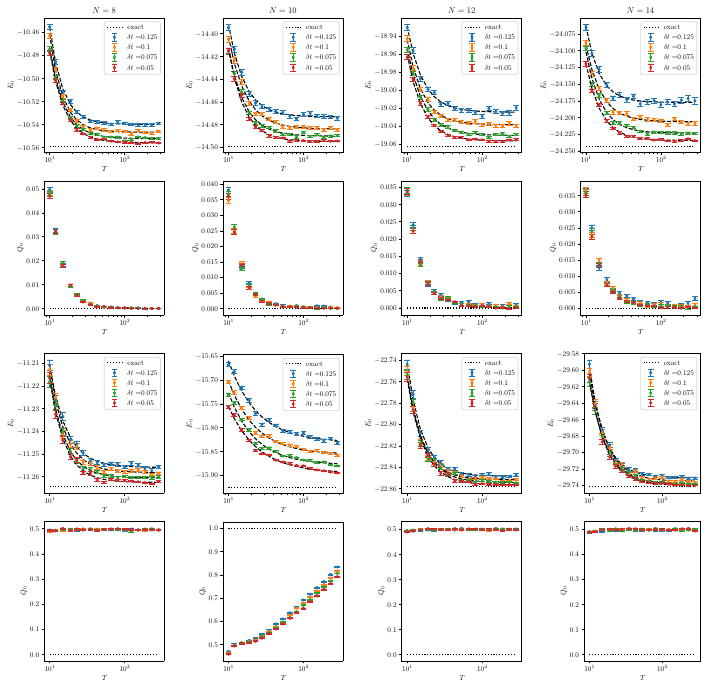}
\caption{Convergence plots of the estimated ground state energy (first and third row) and charge (second and fourth row) for $N=8$ (first column), $N=10$ (second column), $N=12$ (third column) and $N=14$ (fourth column) for two different choices of \QED~parameters. The first two rows refer to a situation where $\Delta$ is large, and there is a nice convergence behaviour. The last two rows show the situations where the gap is closing. Different colors refer to different choices of the Trotter step $\delta t$. The dotted black line is the VEV given by exact diagonalization. The black dashed lines are the results of fitting $T\mapsto E_{0, \text{asymp}} + a T^{-k}$ to the data (for each $\delta t$ separately), and it is clear that $E_{0, \text{asymp}}$ depends on $\delta t$. For all lattices, the data is obtained with $L=4.095$ and in the absence of static charges. The data from the first two rows is obtained with $g=1.6, m=1, \theta=0.4\pi$. The \QED~parameters parameters for the last two rows are: $g=1.8, m=1.5553,\theta=1.2\pi$ for $N=8$ sites,
$g=1.8, m=1.8375,\theta=1.5\pi$ for $N=10$ sites,
$g=1.8, m=2.6842,\theta=1.5\pi$ for $N=12$ sites and 
$g=1.8, m=3.0605,\theta=1.5\pi$ for $N=14$ sites.
}
\label{fig:Tdtconv}
\end{figure*}

In Figure \ref{fig:Tdtconv} we plot the expectation values of $H_{\QEDmat}$ and $Q$ for different choices of $T$ and $\delta t$. 
We consider two distinct situations. One where the \QED~parameters are chosen such that the gap $\Delta$ is large enough for the chosen $T$ to avoid leakage, showing good convergence. In the other case, the \QED~parameters are such that we are closer to where states of different charge cross and so the gap is much smaller.
\par The energy estimations can be fitted to \cite{albashAdiabaticQuantumComputation2018}
\begin{equation}
T\mapsto E_{0, \text{asymp}} + a T^{-k},
\end{equation}
provided that we do so for a fixed $\delta t$. Here $k$ is the convergence rate. For most of the $\delta t$ we considered the convergence rate is $k\approx 2$, whereas for the largest (ie $\delta t=0.125$) the convergence rate is around $k\approx 1.7$, hinting at some residual preasymptotic behaviour (in $\delta t$). The energy estimation $E_{0, \text{asymp}}$ will also be affected by the trotterization error, that is ${E_{0, \text{asymp}}=E_{0, \text{asymp}}(\delta t)}$, and it is possible to obtain an estimation of $E_{0,\text{exact}}$ by fitting
\begin{equation}
\delta t\mapsto E_{0, \text{asymp}}(\delta t) = E_{0} + a' \delta t^{k'}.
\end{equation}
\par Once again we emphasize the importance of monitoring the charge expectation value: in fact the convergence plots of the energy look approximatively the same for both large and small $\Delta$, but in the first case also $Q/g$ is nicely converging to an integer value, whereas in the second case there is barely a convergence, if any. In fact, in the most severe cases the charge VEV remains almost precisely between two integer values. These results hint at a superposition of two states from different charge sectors, signaling thus that the chosen $T$ was not large enough to satisfy the adiabatic condition \eqref{eq:adiabcond}.

\bibliography{apssamp}
\end{document}